\documentclass[sigconf]{acmart}
\AtBeginDocument{%
  }

\setcopyright{acmlicensed}

\copyrightyear{2026}
\acmYear{2026}
\setcopyright{cc}
\setcctype{by}
\acmConference[WWW '26] {Proceedings of the ACM Web Conference 2026}{April 13--17, 2026}{Dubai, United Arab Emirates.}
\acmBooktitle{Proceedings of the ACM Web Conference 2026 (WWW '26), April 13--17, 2026, Dubai, United Arab Emirates}
\acmISBN{979-8-4007-2307-0/2026/04}
\acmDOI{10.1145/3774904.3792573}
\usepackage{xcolor}         
\usepackage{algorithm}       
\usepackage{algorithmic}     
\usepackage{adjustbox}

\usepackage{makecell}
\usepackage{multirow}
\usepackage{array}
\usepackage{amsmath}
\usepackage{amsthm}
\usepackage{booktabs}
\usepackage{graphicx}
\usepackage{bm}
\usepackage{wrapfig}

\usepackage{enumitem}
\usepackage[table]{xcolor}
\usepackage{xspace}
\usepackage{subfigure}

\newcommand{\model}{GFMFR\xspace}

\newcommand{\eg}{\emph{e.g.,}\xspace}
\newcommand{\ie}{\emph{i.e.,}\xspace}

\begin{document}

\title{Multimodal-enhanced Federated Recommendation: A Group-wise Fusion Approach}



\author{Chunxu Zhang}
\authornote{Both authors contributed equally to this research.}
\affiliation{%
  \institution{College of Computer Science and Technology, Jilin University}
  \institution{Key Laboratory of Symbolic Computation and Knowledge Engineering of Ministry of Education, Jilin University}
  \city{Changchun}
  \country{China} \\
  \institution{PolyU Academy for Artificial Intelligence, Hong Kong Polytechnic University}
  \city{Hong Kong}
  \country{China}
  }
\email{zhangchunxu@jlu.edu.cn}

\author{Weipeng Zhang}
\authornotemark[1]
\affiliation{%
   \institution{College of Computer Science and Technology, Jilin University}
  \institution{Key Laboratory of Symbolic Computation and Knowledge Engineering of Ministry of Education, Jilin University}
  \city{Changchun}
  \country{China}
  }
\email{zhangwp24@mails.jlu.edu.cn}

\author{Guodong Long}
\affiliation{%
  \institution{Australian Artificial Intelligence Institute, FEIT, University of Technology Sydney}
  \city{Sydney}
  \country{Australia}
  }
\email{Guodong.Long@uts.edu.au}

\author{Zhiheng Xue}
\affiliation{%
  \institution{College of Computer Science and Technology, Jilin University}
  \institution{Key Laboratory of Symbolic Computation and Knowledge Engineering of Ministry of Education, Jilin University}
  \city{Changchun}
  \country{China}
  }
\email{xuezh24@mails.jlu.edu.cn}

\author{Riting Xia}
\affiliation{%
  \institution{College of Computer Science, Inner Mongolia University}
  \city{Hohhot}
  \country{China}
  }
\email{xiart@imu.edu.cn}

\author{Bo Yang}
\authornote{Corresponding author.}
\affiliation{%
  \institution{College of Computer Science and Technology, Jilin University}
  \institution{Key Laboratory of Symbolic Computation and Knowledge Engineering of Ministry of Education, Jilin University}
  \city{Changchun}
  \country{China}
  }
\email{ybo@jlu.edu.cn}

\renewcommand{\shortauthors}{Chunxu Zhang et al.}

\begin{abstract}
   Federated Recommendation (FR) has emerged as a promising paradigm for addressing the learn-to-rank problem in a privacy-preserving manner. However, effectively incorporating multimodal item features into FR remains an open challenge, due to efficiency constraints, distribution heterogeneity, and feature utilization alignment with the recommendation objective. To tackle these issues, we propose GFMFR, a novel multimodal fusion framework for federated recommendation. Specifically, multimodal representation learning is offloaded to the server, which stores item content and employs a high-capacity encoder to generate expressive representations, thereby alleviating the computational burden on clients. In addition, a group-aware multimodal aggregation mechanism learns shared representations for users with similar interests, enabling knowledge sharing while alleviating distribution heterogeneity. Finally, GFMFR adopts a preference-guided distillation strategy that leverages multimodal information in a way directly aligned with recommendation objectives. The proposed framework can be seamlessly integrated into existing federated recommender systems, enhancing their effectiveness by incorporating multimodal features. Extensive experiments on five benchmark datasets demonstrate that GFMFR consistently outperforms state-of-the-art multimodal FR baselines. The implementation code is available\footnote{\url{https://github.com/Zhangwp2420/GFMFR}}.
\end{abstract}





\maketitle

\section{Introduction}
Driven by rising privacy concerns, Federated Recommendation (FR) has emerged as a prominent paradigm that enables collaborative model training without accessing raw user data, drawing increasing attention~\cite{javeed2023federated,sun2024survey,wang2024horizontal}. Despite this momentum, most existing FR methods still depend heavily on structured item identifiers, such as unique IDs or tabular attributes, to represent user preferences. This limitation becomes particularly pronounced in web-based commerce platforms, where user–item interactions are mediated by multimodal product content such as textual descriptions and images~\cite{liu2024multimodal,malitesta2025formalizing,yang2025curriculum}. Product multimodality is indispensable for capturing fine-grained semantics of items and reflecting user intent at scale, yet it has received limited attention in the FR literature. This gap substantially weakens the applicability of current FR models to realistic Web scenarios~\cite{liu2024alignrec,malitesta2024we,liu2023semantic}.

However, the effective integration of multimodal information into FR remains underexplored, primarily due to several inherent challenges. \textbf{First}, multimodal item attributes often have large raw data sizes and require complex processing for semantic extraction. This imposes substantial storage and computation burdens on resource-constrained client devices. \textbf{Second}, heterogeneous data distributions, particularly differences in modality preferences across clients, make it difficult to learn a unified multimodal representations that generalize well while preserving personalization. \textbf{Third}, aligning multimodal features with recommendation objectives remains a difficult task. Basic fusion methods often introduce redundancy and noise, making it challenging to preserve semantics that are directly relevant for modeling user preferences. These considerations highlight the need for a systematic framework to exploit multimodal knowledge in FR models.

To this end, we introduce a new framework, \textbf{G}roup-wise \textbf{F}usion for \textbf{M}ultimodal-enhanced \textbf{F}ederated \textbf{R}ecommendation (\textbf{\model}). The framework proposes a reallocation of learning responsibilities: Multimodal representation learning takes place centrally on the server side, where comprehensive item content and sufficient computational capacity are accessible. Clients then operate on compact, distilled representations that encapsulate both semantic content and user preference relevance. This design reduces client-side burdens and aligns well with commonly adopted system architectures in real-world deployments (\textbf{Solution for challenge 1}). In addition to improving system efficiency, \model incorporates a group-aware multimodal aggregation mechanism, which enables shared multimodal representation among users in a group with similar preferences. By coordinating learning across users within a group, the mechanism mitigates cross-client heterogeneity while enabling collaborative knowledge propagation (\textbf{Solution for challenge 2}). Moreover, a preference-guided distillation strategy is introduced. It maps multimodal representations into the user preference space and incorporates them into client model optimization via a distillation loss. This approach ensures that multimodal information is utilized in a way that directly aligns with the recommendation objective (\textbf{Solution for challenge 3}). Extensive experiments demonstrate that our method achieves state-of-the-art performance on multimodal recommendation benchmark datasets. Our \textbf{main contributions} are summarized as follows,
\begin{itemize}
    \item We propose \model, a novel conceptual framework that systematically integrates multimodal content into federated recommendation, providing a flexible and effective approach to leverage diverse item modalities.
    \item We design a group-aware multimodal aggregation mechanism to enable fine-grained integration of multimodal representations among users with similar preferences. Additionally, we incorporate a preference-guided distillation loss to align multimodal features with recommendation objectives, balancing knowledge sharing and personalization.
    \item Extensive experiments demonstrate the superior performance of \model compared to competitive baselines. In addition, comprehensive analyses further validate its compatibility, effectiveness and practical applicability.
\end{itemize}

\section{Related Work}
\subsection{Multimodal Recommender System}

With the proliferation of multimedia platforms and diverse content (\eg short videos, images, news), incorporating multimodal information into recommender systems has become crucial for capturing user preferences and item characteristics. Such integration not only enriches representations but also alleviates data sparsity issues~\cite{liu2024multimodal,zhang2024m3oe,zhao2025dvib}, making multimodal recommendation a prominent research direction~\cite{ye2025harnessing,zhuang2025bridging,zhou2025large}. The typical pipeline involves three stages: \textbf{raw feature representation}~\cite{javed2021review,devlin2019bert}, \textbf{feature fusion}~\cite{wang2021dualgnn,tao2022self,yang2025explainable}, and \textbf{recommendation prediction}~\cite{koren2009matrix,meng2025doge}. Raw feature extraction aims to obtain informative modality-specific representations~\cite{chen2022hybrid,he2016deep}; fusion methods~\cite{zhou2023bootstrap,ma2022crosscbr} integrate them into a unified space; and prediction models~\cite{he2016vbpr,xu2025mentor} generate personalized recommendations. Despite their effectiveness, existing centralized approaches raise privacy concerns, limiting their applicability. To address this, we propose a multimodal recommendation framework under federated learning, ensuring privacy protection while retaining recommendation quality.

\subsection{Federated Recommender System}
Federated recommendation (FR) offers personalized services while safeguarding user privacy, making it well-suited for privacy-sensitive scenarios~\cite{alamgir2022federated,chronis2024survey,harasic2024recent,li2024navigating,zhang2025personalized}. Existing FR research spans two major perspectives: \textbf{recommender systems} and \textbf{federated learning}. From the recommender perspective, studies have explored diverse \textbf{model architectures}, including MF~\cite{ammad2019federated,chai2020secure}, MLP~\cite{muhammad2020fedfast,perifanis2022federated}, GNN~\cite{wu2022federated,zhang2024gpfedrec}, and Transformer~\cite{wei2023edge,zhang2025multifaceted}, as well as \textbf{application scenarios} such as cross-domain~\cite{meihan2022fedcdr,zhang2024feddcsr}, cold-start~\cite{zhang2024federated,li2025personalized}, and news recommendation~\cite{qi2020privacy,yi2021efficient}. From the federated learning perspective, efforts have addressed core challenges in \textbf{security}~\cite{liu2022federated,perifanis2023fedpoirec}, \textbf{robustness}~\cite{zhang2022pipattack,ali2024hidattack}, and \textbf{efficiency}~\cite{zhang2023lightfr,NguyenNNHLW24}. However, most FR methods still rely on ID-based item representations, overlooking rich multimodal content (\eg text, images), which can offer valuable semantic cues.

Research on multimodal FR remains limited. Existing methods such as AMMFRS~\cite{feng2024robust} and FedMR~\cite{li2024dynamic} adopt client-side fusion strategies, integrating multimodal features locally: AMMFRS via attention-based weighting and FedMR via interaction-aware fusion modules. However, these approaches assume clients can store and process large-scale multimodal data, which is impractical for resource-constrained devices. FedMMR~\cite{li2024towards} shifts fusion to the server and aligns modalities using a contrastive loss, but focuses primarily on cross-modal alignment while overlooking inter-modal semantic dependencies. To address these limitations, we propose a novel server-side multimodal fusion framework that offloads both storage and representation learning. By performing semantic fusion across modalities and further leveraging collaboration among users with similar preferences, our approach enables more coherent representation learning and fosters collaborative knowledge sharing in FR systems.

\section{Preliminary}
\paragraph{General FR framework.}
Let $\mathcal{U}$ and $\mathcal{I}$ denote the user set and item set in the recommender system, and $N=|\mathcal{U}|$ and $M=|\mathcal{I}|$ are the total number of users and items respectively. For a recommendation model $\mathcal{F}$ parameterized by $\theta$, it is generally composed of three modules, \ie user embedding $E_{\mathcal{U}}$, item embedding $E_{\mathcal{I}}$ and prediction function $P$. Given user $u$ and item $i$, the recommendation model makes rating predictions as $\hat{\mathcal{R}}_{ui}=\mathcal{F}(u, i|\theta)$. In the federated recommender system, each user $u$ acts as a client who holds the personal data $\mathcal{R}_u$, including all the historically rated items. To implement a FR model, the system alternates between the clients and server and performs the following steps: 

\textit{(1) Client-side model training:} The client $u$ trains the locally maintained recommendation model $\mathcal{F}_\theta$ with personal data $\mathcal{R}_u$. \textit{(2) Server-side model aggregation:} A server collects the updated models from clients and performs a global aggregation operation to achieve a unified model. \textit{(3) Global model synchronization:} The unified model would be sent to the clients to improve their models' training in the next round. Formally, the optimization objective of FR can be formulated as follows,
\begin{equation}\label{eq:ob_FR}
    \min_{\{E_{\mathcal{U}}^u\}_{u=1}^N, E_{\mathcal{I}}, P} \sum_{u=1}^N \alpha_{u} \mathcal{L}_{rec}(\theta; \mathcal{R}_u)
\end{equation}
where $\mathcal{L}_{rec}$ is local recommendation model loss of the $u$-th client, and $\alpha_u$ is the weight of the $u$-th client during global aggregation. For example, $\alpha_u$ is usually proportional to the client's personal data volume and the total system data volume~\cite{mcmahan2017communication}. In existing FR research, the user embedding module $E_{\mathcal{U}}$ is typically treated as a private component and retained locally on each client, while the item embedding $E_{\mathcal{I}}$ and prediction function $P$ are shared with the server for global collaborative training.

\section{Group-wise Fusion for Multimodal-enhanced FR (\model)}
In this section, we introduce the proposed Group-wise Fusion for Multimodal-enhanced FR (\model). The framework delegates multimodal representation learning to the server, aligning with practical deployments where intensive computations are handled centrally. To enhance cross-client knowledge sharing, we design a group-aware aggregation mechanism enabling shared representation among users with similar preferences. A preference-guided distillation strategy further maps multimodal features into the preference space and distills them into local models, enabling effective integration of multimodal signals into FR. The following subsections detail the framework’s components.
\begin{figure*}[!t]
    \centering
    \includegraphics[width=1.0\textwidth]{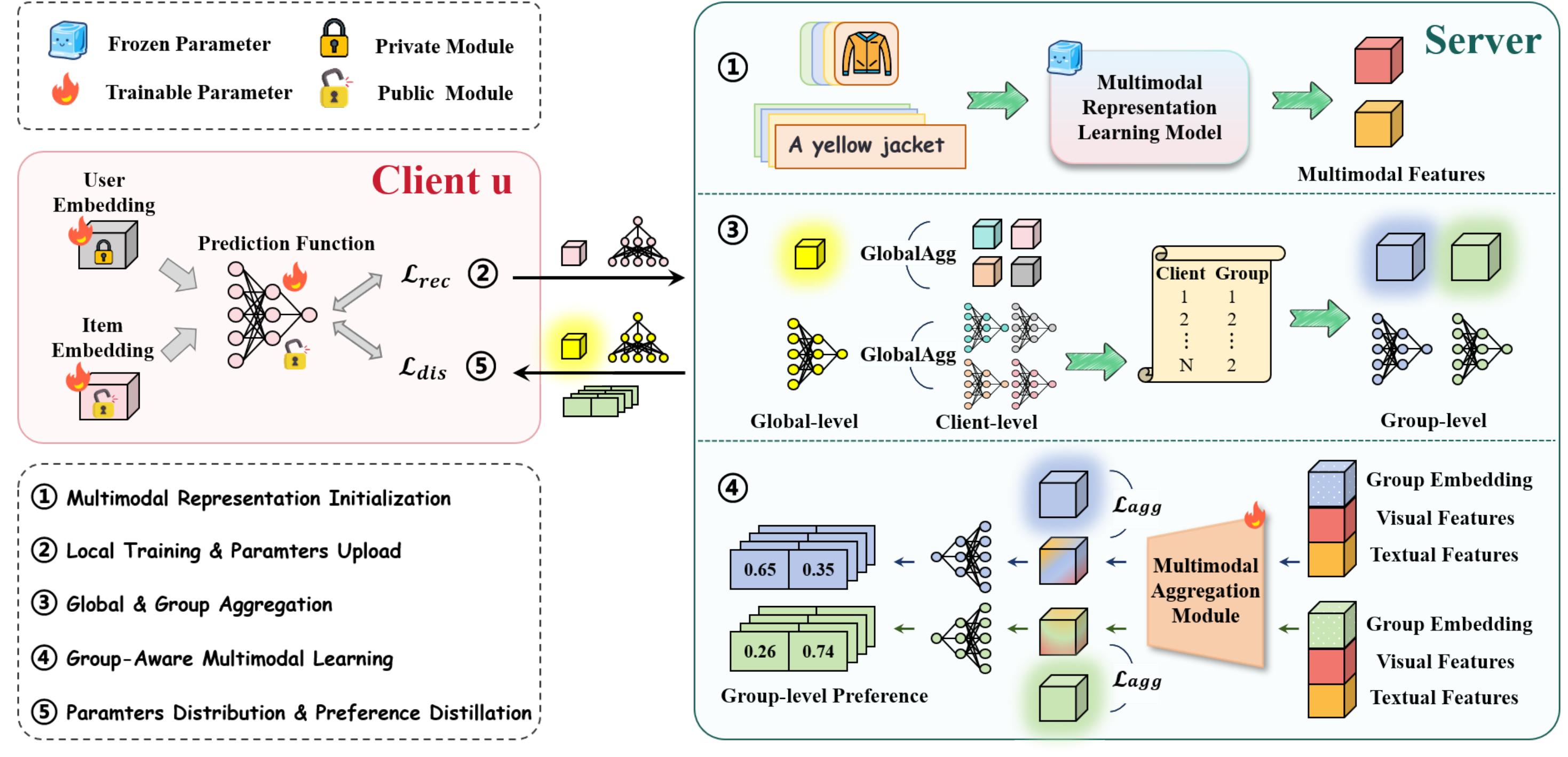}
   
    \caption{Overview of the proposed \model. The entire process unfolds in five stages. Initially, before the federated optimization begins, the server generates initial multimodal representations based on raw multimodal data (Step \textcircled{1}). During each communication round, clients first train local recommendation models using their private data and upload the resulting item embeddings and prediction functions to the server (Step \textcircled{2}). The server then performs global aggregation on the received embeddings and functions, groups users based on their prediction functions, and computes group-level item embeddings and prediction functions (Step \textcircled{3}). Leveraging these group-level embeddings, the server trains a multimodal aggregation model and projects the fused representations into the preference space via the group-level prediction functions (Step \textcircled{4}). Finally, the global parameters and group-level preference representations are sent back to the clients to refine their local models (Step \textcircled{5}).}
    \label{fig:framework}
\end{figure*}

\subsection{Multimodal Representation Initialization}
In multimodal recommendation, item profiles are enriched with various signals such as textual descriptions and product images. To make these diverse sources usable, raw inputs from each modality should first be processed by modality-specific encoders, ensuring that their distinct characteristics are preserved in model-compatible form. Handling these inputs on clients is often infeasible due to limited computational and storage resources, whereas the server is equipped with abundant computational capacity and can leverage large-scale pre-trained backbones to generate semantically rich and well-structured representations. To utilize these signals effectively, this work relies on the server to learn initial embeddings directly from the raw data using a multimodal representation learning model, producing initial embeddings \(\mathcal{E}_m \in \mathbb{R}^{M \times d_1}\) for \(m=1,\ldots,k\), where $k$ is the number of item
modalities and \(d_1\) denotes the embedding dimension, thereby forming a solid foundation for subsequent multimodal fusion and integration of the representations in FR systems.

\subsection{Group-Aware Multimodal Aggregation}
In FR systems, users exhibit substantial heterogeneity in how they rely on different modalities. For instance, younger users may focus more on visual cues, whereas older users may place greater emphasis on textual descriptions. This diversity means that a single global multimodal representation cannot accommodate all users, while learning fully user-specific fusion quickly becomes inefficient and unstable. To better balance generalization and personalization, we propose a group-aware multimodal aggregation mechanism that learns shared representations for preference-aligned user groups, capturing intra-group regularities and enabling more effective multimodal knowledge transfer.

\subsubsection{Cluster clients into groups.}
To effectively capture shared patterns among users with similar preferences, we begin by organizing clients into groups, which forms the basis for learning group-level representations. This grouping is based on clients' prediction functions, which reflect how users make decisions over items and thus implicitly encode their preference patterns. Instead of directly sharing raw interaction or feature data, each client transmits its learned prediction function $\{P_u\}_{u=1}^N$, preserving privacy while still providing meaningful signals for grouping. The server then collects the set of functions $\{P_u\}_{u=1}^N$ and applies a clustering algorithm to partition users into groups with similar decision behaviors, formally expressed as
\begin{equation}\label{eq:cluster}
    G_1, G_2, ..., G_g = \text{ClusterAlgorithm}(\{P_u\}_{u=1}^N)
\end{equation}
where $g$ denotes the total number of user groups. Standard clustering methods such as K-Means or hierarchical clustering can be applied, and the resulting groups serve as the basis for learning shared group-level multimodal representations.

\subsubsection{Learn group-level multimodal features.}
Once the user groups $\{G_1, ..., G_g\}$ have been identified, we learn shared multimodal representations within each group. Given the initial item representations $\{\mathcal{E}_m\}_{m=1}^k$ produced by the server, a multimodal aggregation module integrates these features for each group, producing group-level embeddings that consolidate the multimodal signals of items with respect to the user cluster. 

The aggregation module consists of a learnable group embedding $E_G \in \mathbb{R}^{g \times d_2}$, which assigns a unique vector representation to each user group, and an aggregation function $S_\phi$ that combines item-level multimodal embeddings with the group embedding. Formally, for each group $l=1,\dots,g$, the aggregation is formulated as,
\begin{equation}\label{eq:aggregation}
    Q_l = S_\phi(\{\mathcal{E}_m\}_{m=1}^k, E_G(l))
\end{equation}
where $E_G(l)$ is the embedding for group $l$, and $Q_l$ denotes the aggregated multimodal features.

To optimize this module, we leverage group-level item embeddings as supervisory signals, guiding the aggregated multimodal features to better reflect the collective item representations within each user group. Specifically, for each group $l$, the group item embedding $C_l$ is obtained by aggregating the local embeddings of its members, \eg $C_l = \text{AvgPooling}(\{E_\mathcal{I}^u\}_{u \in G_l})$. These embeddings summarize how items are perceived by users within the group, providing a target for group-level aggregation. The parameters of the aggregation function $S_\phi$ and the group embeddings $E_G$ are then updated by minimizing,
\begin{equation}\label{eq:ob_aggregation}
    \min_{E_G, \phi} \frac{1}{g} \sum_{l=1}^g \mathcal{L}_{agg}(E_G, \phi; C_l)
\end{equation}
where $\mathcal{L}_{agg}$ denotes the aggregation loss, typically instantiated as the mean squared error between the aggregated group embedding $Q_l$ and the corresponding group item embedding $C_l$. This optimization encourages the group-level multimodal representations to capture consistent cross-modal semantics while reflecting the collective perception of items within each user group, facilitating more coherent and informative embeddings for subsequent recommendation learning.

\subsection{Preference-Guided Distillation}
To effectively leverage multimodal signals, we propose a preference-guided distillation strategy. This approach projects group-level multimodal features into the user preference space, providing auxiliary supervision during client-side training. In this way, clients incorporate multimodal signals directly aligned with recommendation objectives, enhancing the learning of preference-relevant representations and enabling more personalized recommendations. 

\subsubsection{Map group-level multimodal features into preference space.}
After obtaining group-level multimodal features $\{Q_l\}_{l=1}^g$, the next step is to project them into the preference space to make them usable for client-side training. This projection allows the multimodal representations, which encode rich item information, to be aligned with how users express preferences in each group. Specifically, we obtain the group prediction function by averaging client functions within each group, $\overline{P}_l = \text{AvgPooling}(\{P_u\}_{u \in G_l})$. This function captures the overall decision patterns of users in the group, reflecting their shared tendencies in ranking and evaluating items. We then apply $\overline{P}_l$ to the group-level multimodal features $Q_l$ to generate the group’s preference-aligned representation,
\begin{equation}\label{eq:preference_prediction}
    E_{\mathcal{M}\_l} = \overline{P}_l(Q_l)
\end{equation}
Once the group's preference representations $\{E_{\mathcal{M}\_l}\}_{l=1}^g$ are obtained, the server will distribute them to the clients to improve local model's training.

\subsubsection{Distill group preference to client recommendation models.}
Each client trains its recommendation model locally on private data. Upon receiving the group-level preference representation from the server, it incorporates this as auxiliary supervision to refine the local model. This distillation guides the client model to align its predictions with group-level user preferences, thereby directly shaping the learning of recommendation-relevant behaviors rather than merely fusing features. Formally, the distillation objective for client $u$ in group $l$ is,
\begin{equation}\label{eq:ob_distillation}
    \min_{\theta_u} \mathcal{L}_{dis}(\theta_u; E_{\mathcal{M}\_l})
\end{equation}
where $\theta_u$ denote the $u$-th client's model parameters, and $\mathcal{L}_{dis}$ is the distillation loss function, \eg the KL divergency between the client’s predicted outputs $\hat{\mathcal{R}}_u$ and the group-level preference representation $E_{\mathcal{M}\_l}$. By leveraging these group-level representations, client models incorporate multimodal signals consistently with group tendencies, facilitating knowledge transfer across users with similar preferences while maintaining computational efficiency.

\begin{algorithm}[htbp]
    \begin{minipage}{1.0\linewidth}
    \caption{Group-wise Fusion for Multimodal-enhanced Federated Recommendation}
    \label{algorithm}
    \hspace*{0.05in}{\bf ServerProcedure}:
    \begin{algorithmic}[1]
        
        \STATE Initialize globally shared item embedding $E_{\mathcal{I}}^{(1)}$ and prediction function $P^{(1)}$
        \STATE Obtain initial multimodal representations $\{\mathcal{E}_m\}_{m=1}^k$ with a powerful model
   
        \FOR{each round $t=1, 2, \dots, T$}
            \STATE Randomly select a client subset $B^{(t)}$ from $N$ clients based on sampling ratio $\beta$

        \FOR{client $u \in B^{(t)}$ \textbf{in parallel}}
        \STATE $E_{\mathcal{I}}^u$,
                $P_u$ $\leftarrow$ ClientUpdate($u, t, E_{\mathcal{I}}^{(t)}, P^{(t)}$) 
        \ENDFOR    
        
        \STATE $E_{\mathcal{I}}^{(t+1)}, P^{(t+1)} \leftarrow \text{GlobalAgg}(\{E_{\mathcal{I}}^u\}_{u \in B^{(t)}}, \{P_u\}_{u \in B^{(t)}})$; 
        
        \STATE Cluster clients into groups $\{G_1, G_2, ..., G_g\}^{(t)}$ based on $\{P_u\}_{u \in B^{(t)}}$ with Eq. (\ref{eq:cluster})
        
        \STATE Train a multimodal aggregation module with group-specific item embedding with Eq. (\ref{eq:ob_aggregation})
        
        \STATE Obtain the group-level multimodal representations $\{Q_l^{(t)}\}_{l=1}^g$ with Eq. (\ref{eq:aggregation})
        
        \STATE Calculate the group preference representations $\{E_{\mathcal{M}\_l}^{(t)}\}_{l=1}^g$ via the group-specific prediction function with Eq. (\ref{eq:preference_prediction})

        \STATE Distribute $\{E_{\mathcal{M}\_l}^{(t)}\}_{l=1}^g$ to client $i \in B^{(t)}$ to refine local model with Eq. (\ref{eq:ob_distillation}). (with $\lambda$ balancing the contribution of distillation loss against the local recommendation loss)
        \ENDFOR
        
    \end{algorithmic}

    \hspace*{0.05in}{\bf ClientUpdate($u, t, E_{\mathcal{I}}^{(t)}, P^{(t)}$)}:
    \begin{algorithmic}[1]
    
         \IF{$t = 1$}
            \STATE Initialize user embedding $E_{\mathcal{U}}$
         \ELSE
            \STATE Load user embedding from the last round
        \ENDIF
         
        \STATE Initialize $(E_{\mathcal{I}}, P)$ with $(E_{\mathcal{I}}^{(t)}, P^{(t)})$
                
        \FOR{$h = 1, 2, \dots, H$} 
            \STATE Update $E_{\mathcal{U}}, E_{\mathcal{I}}, P$ based on personal data $\mathcal{R}_u$ with recommendation loss $\mathcal{L}_{rec}$
        \ENDFOR

        \STATE \bf Return $E_{\mathcal{I}}, P$
    \end{algorithmic}
    \end{minipage}
\end{algorithm}

\subsection{Optimization Objective and Algorithm}

\subsubsection{Optimization objective.}
The goal of \model is to jointly learn group-level multimodal representations and guide client models with group-aligned preference signals in the FR framework. We formulate the overall optimization objective as follows,
\begin{gather}\label{eq:ob_overall}
    \min_{\theta} \sum_{u=1}^N \alpha_{u} \left( \mathcal{L}_{rec}(\theta; \mathcal{R}_u) + \lambda \cdot \mathcal{L}_{dis}(\theta; E_{\mathcal{M}\_l}) \right) \\
    \text{s.t.} \quad (E_G^*, \phi^*) = \arg\min_{E_G, \phi} \frac{1}{g} \sum_{l=1}^g \mathcal{L}_{agg}(E_G, \phi; C_l),  \notag \\ \quad Q_l = S_{\phi^*}(\{\mathcal{E}_m\}_{m=1}^k, E_G^*(l)), \quad E_{\mathcal{M}\_l} = \overline{P}_l(Q_l) \notag
\end{gather}
where $\lambda$ is the coefficient of distillation loss. $(E_G^*, \phi^*)$ are the optimized parameters of the multimodal aggregation module, used to generate group-level multimodal representations $Q_l$, which are subsequently mapped to preference representations $E_{\mathcal{M}\_l}$. Here, $l$ denotes the group index assigned to the $u$-th client, which is dynamically updated during the federated optimization process.

\begin{table*}[!t]
\centering
\caption{Overall performance comparison of our proposed \model and baselines integrated into three representative FR backbone models. ``H@50'' and ``N@50'' are abbreviations for HR@50 and NDCG@50. Bold indicates the best result, underline marks the second best result, and ``*'' denotes a statistically significant improvement over the best baseline (two-sided t-test, p < 0.05).}
\label{tab:overall_performance}
\resizebox{\textwidth}{!}{
\begin{tabular}{l|cc|cc|cc|cc|cc}
\toprule[1.0pt]
\multirow{2}{*}{\textbf{Model}} & \multicolumn{2}{c|}{\textbf{Beauty}} & \multicolumn{2}{c|}{\textbf{Tools\_and\_Home}} & \multicolumn{2}{c|}{\textbf{Toys\_and\_Games}} & \multicolumn{2}{c|}{\textbf{Digital\_Music}} & \multicolumn{2}{c}{\textbf{Office\_Products}} \\
& H@50 & N@50 & H@50 & N@50 & H@50 & N@50 & H@50 & N@50 & H@50 & N@50 \\
\midrule[0.2pt]
\textbf{FedNCF} & 0.1450 & 0.0407 & 0.1893 & 0.0483 &\underline{0.2858} & \underline{0.2436} & \underline{0.2249} & \underline{0.1822} & 0.0015 & 0.0004 \\
\textbf{w/ FedMMR} & 0.1331 & 0.0358 & 0.1198 & 0.0287 & 0.0354 & 0.0089 & 0.0200 & 0.0052 & 0.0130 & 0.0035 \\
\textbf{w/ AMMFRS} & 0.1549 & \underline{0.0449} & \underline{0.1947} & \underline{0.0503} & 0.0733 & 0.0221 & 0.0417 & 0.0098 & \underline{0.0145} & \underline{0.0039} \\
\textbf{w/ FedMR} & \underline{0.1601} & 0.0378 & 0.1777 & 0.0405 & 0.0565 & 0.0141 & 0.0267 & 0.0074 & 0.0142 & 0.0036 \\
\textbf{w/ FedMLP} & 0.1537 & 0.0364 & 0.1653 & 0.0444 & 0.0558 & 0.0120 & 0.0222 & 0.0053 & 0.0039 & 0.0011 \\
\cellcolor{blue!20}\textbf{w/ \model} & \cellcolor{blue!20}\textbf{0.2067*} & \cellcolor{blue!20}\textbf{0.0552*} & \cellcolor{blue!20}\textbf{0.2062*} & \cellcolor{blue!20}\textbf{0.0527*} & \cellcolor{blue!20}\textbf{0.3072*} & \cellcolor{blue!20}\textbf{0.2624*} & \cellcolor{blue!20}\textbf{0.2521*} & \cellcolor{blue!20}\textbf{0.2238*} & \cellcolor{blue!20}\textbf{0.0255*} & \cellcolor{blue!20}\textbf{0.0070*} \\
\midrule[1.0pt]
\textbf{PFedRec} & 0.1427 & 0.0320 & 0.1596 & 0.0443 & \underline{0.1412} & \underline{0.0626} & \underline{0.1287} & \underline{0.0728} & 0.0111 & 0.0027 \\
\textbf{w/ FedMMR} & 0.1383 & 0.0351 & 0.1212 & 0.0301 & 0.0401 & 0.0097 & 0.0228 & 0.0054 & 0.0118 & 0.0031 \\
\textbf{w/ AMMFRS} & 0.1594 & 0.0418 & \underline{0.1791} & 0.0451 & 0.0829 & 0.0258 & 0.0455 & 0.0142 & 0.0139 & 0.0036  \\
\textbf{w/ FedMR} & \underline{0.2022} & \underline{0.0545} & 0.1756 & \underline{0.0600} & 0.0524 & 0.0205 & 0.0612 & 0.0321 & 0.0164 & 0.0046\\
\textbf{w/ FedMLP} & 0.1537 & 0.0361 & 0.1645 & 0.0417 & 0.1207 & 0.0516 & 0.0875 & 0.0361 & \underline{0.0266} & \underline{0.0074} \\
\cellcolor{blue!20}\textbf{w/ \model} & \cellcolor{blue!20}\textbf{0.2024*} & \cellcolor{blue!20}\textbf{0.0730*} & \cellcolor{blue!20}\textbf{0.2269*} & \cellcolor{blue!20}\textbf{0.0612*} & \cellcolor{blue!20}\textbf{0.2979*} & \cellcolor{blue!20}\textbf{0.2613*} & \cellcolor{blue!20}\textbf{0.2943*} & \cellcolor{blue!20}\textbf{0.2820*} & \cellcolor{blue!20}\textbf{0.0614*} & \cellcolor{blue!20}\textbf{0.0198*} \\
\midrule[1.0pt]
\textbf{FedRAP} & 0.1423 & 0.0363 & 0.1641 & 0.0560 & \underline{0.0616} & \underline{0.0226} & 0.0531 & 0.0273 & 0.0118 & 0.0033 \\
\textbf{w/ FedMMR} & 0.1331 & 0.0374 & 0.1291 & 0.0324 & 0.0397 & 0.0099 & 0.0224 & 0.0057 & 0.0103 & 0.0028 \\
\textbf{w/ AMMFRS} & 0.1522 & 0.0421 & 0.1384 & 0.0342 & 0.0414 & 0.0102&
0.0182 & 0.0043 & 0.0160 & 0.0040\\
\textbf{w/ FedMR} & 0.1568 & 0.0366 & \underline{0.1896} & \underline{0.0578} & 0.0480 & 0.0123 & 0.0280 & 0.0075 & 0.0112 & 0.0028\\
\textbf{w/ FedMLP} & \underline{0.1771} & \underline{0.0479} & 0.1624 & 0.0562 & 0.0547 & 0.0225 & \underline{0.0555} & \underline{0.0298} & \underline{0.0178} & \underline{0.0051} \\
\cellcolor{blue!20}\textbf{w/ \model} & \cellcolor{blue!20}\textbf{0.1949*} & \cellcolor{blue!20}\textbf{0.0525*} & \cellcolor{blue!20}\textbf{0.2587*} & \cellcolor{blue!20}\textbf{0.0606*} & \cellcolor{blue!20}\textbf{0.2730*} & \cellcolor{blue!20}\textbf{0.1978*} & \cellcolor{blue!20}\textbf{0.2852*} & \cellcolor{blue!20}\textbf{0.1845*} & \cellcolor{blue!20}\textbf{0.0647*} & \cellcolor{blue!20}\textbf{0.0208*} \\
\bottomrule[1.0pt]
\end{tabular}
}
\end{table*}

\subsubsection{Algorithm.}
To solve the above optimization objective, we conduct an iterative algorithm between the server and clients to learn the multimodal FR system. As illustrated in Algorithm~\ref{algorithm}, \textbf{before the federated optimization begins}, the server first performs a unified initialization of the globally shared parameters, including the item embedding $E_{\mathcal{I}}$ and the prediction function $P$. In parallel, high-quality initial multimodal representations $\{\mathcal{E}_m\}_{m=1}^k$ are obtained using a powerful multimodal representation learning model. \textbf{In each communication round $\bm{t}$}, selected clients $u \in B^{(t)}$ initialize local models with the latest shared parameters $E_{\mathcal{I}}^{(t)}$ and $P^{(t)}$, perform local updates using personal data $\mathcal{R}_u$, and upload the updated shared components to the server. Upon receiving the updated shared parameters $\{E_{\mathcal{I}}^u\}_{u \in B^{(t)}}$ and $\{P_u\}_{u \in B^{(t)}}$, the server performs global aggregation to obtain the new shared parameters $E_{\mathcal{I}}^{(t+1)}$ and $P^{(t+1)}$. Concurrently, it clusters clients into groups $\{G_1, G_2, ..., G_g\}^{(t)}$ based on the shared prediction functions $\{P_u\}_{u \in B^{(t)}}$, and trains a multimodal aggregation module with group-specific item embeddings $\{C_l^{(t)}\}_{l=1}^g$ to derive group-level multimodal representations $\{Q_l^{(t)}\}_{l=1}^g$. These representations are then transformed into group preference signals $\{E_{\mathcal{M}\_l}^{(t)}\}_{l=1}^g$ via the group-specific prediction function $\{\overline{P}_l^{(t)}\}_{l=1}^g$, which are subsequently distributed to corresponding clients $u \in B^{(t)}$ to refine their local models with the distillation loss $\mathcal{L}_{dis}$.

\subsection{Discussion about Practical Applicability}
\subsubsection{Usage guideline.}
The proposed multimodal-enhanced FR framework emphasizes modularity and compatibility for seamless integration into existing FR pipelines. The multimodal representation learning is decomposed into three stages for practical deployment and customization. First, initial modality-specific representations can be obtained by using powerful pre-trained models on the server or trained from scratch, supporting flexible domain adaptation. Second, the group-aware multimodal aggregation module supports flexible grouping strategies (\eg behavior, geography) and aggregation mechanisms (\eg MLP~\cite{rosenblatt1958perceptron}, self-attention~\cite{vaswani2017attention}, mixture-of-experts~\cite{jacobs1991adaptive}), with adjustable aggregation frequency balancing personalization and efficiency. Third, the preference-guided distillation can be replaced by alternative integration methods (\eg adapter modules, attention fusion) to align multimodal knowledge with local objectives. These design choices ensure adaptability across diverse system constraints and scenarios.

\subsubsection{Efficiency analysis.}
Multimodal item attributes, derived from raw data such as images, text, and audio, require high-capacity models for effective encoding. In \model, storing and encoding these attributes is fully offloaded to the server, leveraging its superior computational and storage resources. This eliminates the burden on resource-limited clients from handling large multimodal data or complex models. The server further compresses rich multimodal semantics into low-dimensional preference representations (\eg dimension 2 in implicit feedback tasks), which are then transmitted to clients to enhance local training. This approach preserves task-relevant semantics while significantly reducing communication overhead in federated optimization.

\subsubsection{Security analysis.}
Unlike conventional ID-based representations, multimodal features are semantically rich but may inadvertently encode sensitive user attributes when learned locally, increasing the risk of privacy breaches via gradient inversion or attribute inference. In our framework, representation learning is entirely server-side, and clients receive only compact item preference vectors derived from multimodal features, which are less informative and more resilient to such attacks. This design mitigates privacy risks associated with multimodal signals in FR while remaining compatible with ID-based architectures, and amenable to integration with techniques like Differential Privacy~\cite{choi2018guaranteeing} and Homomorphic Encryption~\cite{acar2018survey}. See experiments for further analysis.

\section{Experiment}
\subsection{Experimental Setup}
\paragraph{Dataset.}
We evaluate our method on five multimodal recommendation datasets from Amazon \cite{hou2024bridging}, including Beauty, Tools\_and\_Home (TH), Toys\_and\_Games(TG), Digital\_Music(DM), Office\_Products(OP). Detailed data descriptions are summarized in Appendix \ref{app:dataset}. 

\paragraph{Baselines.}
Our proposed \model \space is a general multimodal FR framework that can be seamlessly integrated with existing ID-based FR models. To verify its versatility, we select three advanced ID-based FR models as the backbones, including FedNCF~\cite{perifanis2022federated}, PFedRec~\cite{zhang2023dual} and FedRAP~\cite{li2024federated}. In addition, we include three existing multimodal FR models as baselines for comparison, including FedMMR\cite{li2024towards}, AMMFRS\cite{feng2024robust} and FedMR\cite{li2024dynamic}. We also extend a centralized multimodal recommendation model~\cite{zhang2017joint} to the federated setting (named FedMLP) by employing an MLP-based fusion strategy on the client side, thereby enriching the comparison pool with a comprehensive coverage of commonly used multimodal fusion techniques. More details can be found in Appendix \ref{app:baseline}.


\paragraph{Evaluation protocol and implementation details.}
 We evaluate the performance using two ranking-based metrics for top-k recommendation, \ie Hit Rate (HR@k) and Normalized Discounted Cumulative Gain (NDCG@k). Details can be found in Appendix \ref{app:implementation}.

 \begin{figure}[tbp]
    \includegraphics[width=\linewidth]{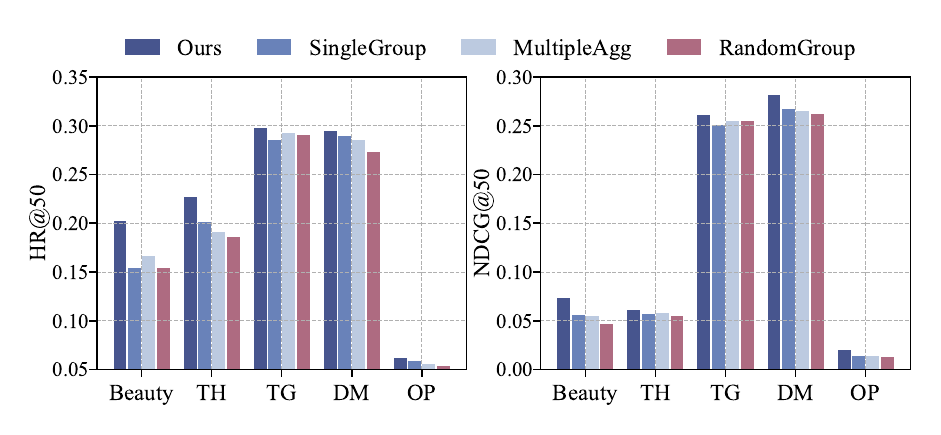}
    \caption{Ablation Study of Group-aware Multimodal Aggregation Mechanism.}
    \label{fig:ablation_1}
\end{figure}

\begin{figure}[tbp]
    \includegraphics[width=\linewidth]{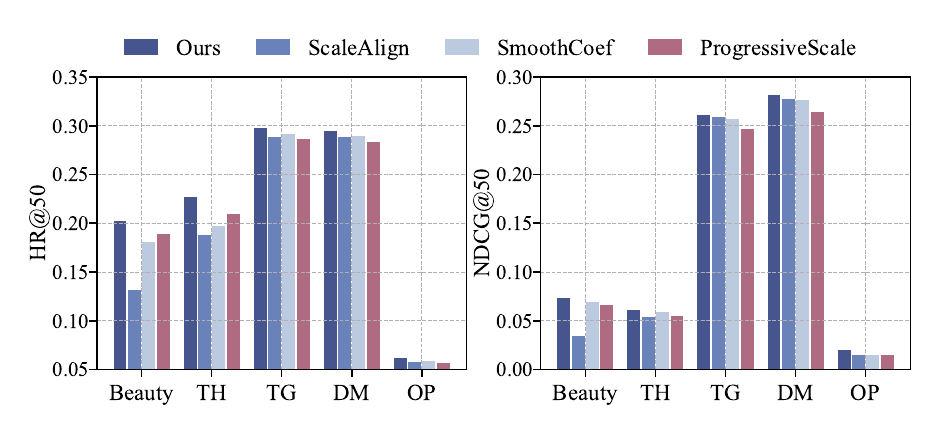}
    \caption{Ablation Study of Preference-guided Distillation Strategy.}
    \label{fig:ablation_2}
\end{figure}

\subsection{Overall Comparison}\label{main_results}
\paragraph{Overall performance.}
Table~\ref{tab:overall_performance} reports HR@50 and NDCG@50 comparisons across five datasets. Key insights from the results are as follows: \textit{\textbf{(1) Integrating multimodal information into FR models can typically enhance system performance.}} Effective integration of semantic cues in multimodal data is crucial for the system to achieve a deeper understanding of items. This deeper understanding allows for a more accurate characterization of items, which in turn enhances system performance. \textit{\textbf{(2) Our method demonstrates superior performance compared to baselines.}} Existing multimodal FR methods primarily generate a unified item representation by leveraging specific fusion strategies to enhance features. In contrast, our method leverages multimodal information to learn preference representations that are more aligned with the optimization objectives of recommendation models, leading to improved performance. \textit{\textbf{(3) Our method showcases exceptional compatibility across various FR backbones.}} \model serves as a general multimodal FR framework that can be seamlessly integrated with existing ID-based methods. Experimental results demonstrate that incorporating our model into three different FR backbones consistently yields the best performance across all datasets, highlighting its strong compatibility.

\paragraph{Efficiency comparison.}
We further evaluate and compare the efficiency of our method and baselines from three aspects: (1) client-side storage overhead, including multimodal data representations (preferences) and recommendation model parameters; (2) communication overhead, covering the initial distribution of multimodal representations and the per-round transmission cost; and (3) per-epoch training time of the client-side recommendation model. Our method consistently achieves lower storage and communication costs as well as faster per-epoch training, demonstrating superior overall efficiency. More details can be found in Appendix \ref{app:efficiency}.


\subsection{Ablation Study}
This subsection examines the impact of the group-aware multimodal aggregation mechanism and the preference-guided distillation strategy on model performance. The analysis clarifies their individual contributions and demonstrates how each component enhances overall effectiveness. We evaluate our method by incorporating it into PFedRec and report the HR@50 and NDCG@50 results across all datasets.

\paragraph{Group-aware multimodal aggregation mechanism.}
We perform a step-wise ablation study with three key variants: replace preference-based clustering with random grouping to assess grouping quality (\textbf{RandomGroup}); treat all users as one group to examine the need for group differentiation (\textbf{SingleGroup}); and use group-specific aggregation modules instead of a shared one to evaluate the benefit of shared representation learning (\textbf{MultipleAgg}). As shown in Figure~\ref{fig:ablation_1}, random grouping notably degrades performance, highlighting the importance of behavior-aware user partitioning. Treating all users as a single group also impairs results, as it overlooks individual preferences. Furthermore, using multiple aggregation modules reduces performance due to increased complexity and the absence of shared representation learning, compromising generalization.

\paragraph{Preference-guided distillation strategy.}
The preference-guided distillation strategy leverages group-level preference representations to refine client training. To examine the effect of preference signal utilization, we compare four scheduling strategies for the coefficient $\lambda$ (as defined in Eq. \ref{eq:ob_overall}) of the distillation loss: (1) match the scale of recommendation loss (\textbf{ScaleAlign}), (2) smooth with the previous round’s coefficient (\textbf{SmoothCoef}), (3) progressively increase the coefficient (\textbf{ProgressiveScale}), and (4) our strategy combines (2) and (3) (\textbf{Ours}). As shown in Figure~\ref{fig:ablation_2}, our method achieves the best performance by effectively balancing training stability and the progressive integration of user preferences.

\begin{figure}[tbp]
    \centering
    \includegraphics[width=\linewidth]{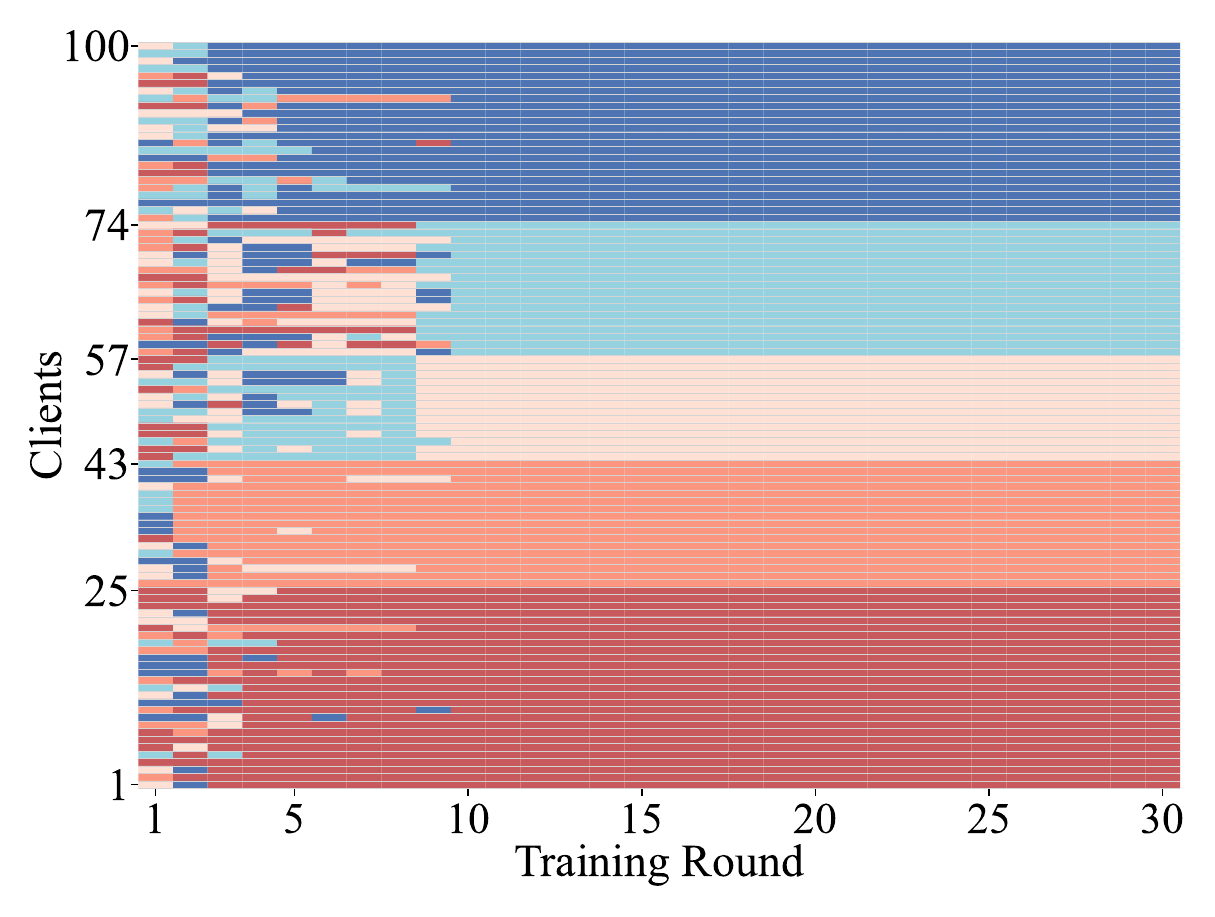}
    \caption{Visualization of client grouping dynamics during training on dataset Tools\_and\_Home. Each color indicates a distinct group.}
    \label{fig:stable_group} 

\end{figure}

\subsection{Hyper-Parameter Analysis}
This subsection investigates the impact of several key hyperparameters on model performance, including the number of user groups $g$, the number of local training epochs per client $H$, and the client sampling ratio per round $\beta$. \textbf{For the number of user groups $\bm{g}$}, experimental results reveal that model performance is sensitive to the number of user groups, with the best performance varying across datasets. An appropriate group number enables effective user partitioning and better captures preference diversity. \textbf{For the number of local training epochs $\bm{H}$}, setting local epochs to 5 generally yields the best performance. Too few local epochs result in underfitted and noisy updates, whereas excessive training can induce overfitting and escalate computational costs. \textbf{For the client sampling ratio $\bm{\beta}$}, model performance remains generally comparable across different sampling ratios. Lower ratios may lead to slower convergence, whereas higher ratios help accelerate training and improve resource efficiency. Complete results and analysis on all datasets can be found in Appendix \ref{app:hyperparameter}.

\subsection{Visualization Analysis}
To better understand how our model organizes users into groups, we visualize the grouping dynamics during training on dataset Tools\_and\_Home. We randomly sample 100 users and track their group assignments across training rounds, with different colors indicating different groups. As shown in Figure \ref{fig:stable_group}, user groupings fluctuate frequently in early stages but gradually stabilize as training progresses. This indicates that the model captures latent user affinities, clustering users with similar characteristics into consistent groups. These results align with real-world scenarios, where different user cohorts exhibit distinct multimodal preferences, and support our motivation of leveraging group-level preferences to enhance the role of multimodal signals in recommendation.

\subsection{Privacy Protection-Enhanced \model}
To further enhance the privacy protection of the proposed model, we integrate local differential privacy (LDP) by perturbing all shared parameters with Laplacian noise prior to transmission to the server. Specifically, we experiment with five levels of noise intensity by setting the noise scale $\delta \in \{1, 2, 3, 4, 5\}$ under unit sensitivity, which corresponds to privacy budgets $\epsilon \in \{1.0, 0.5, 0.33, 0.25, 0.2\}$. As shown in  Table \ref{tab:performance_of_ldp}, increasing the noise intensity improves privacy protection but leads to a gradual decline in model accuracy. In practical deployments, a moderate noise level can serve as a balanced trade-off, providing sufficient privacy without significantly sacrificing recommendation quality.

\begin{table}[!t]
\centering
\caption{Performance of integrating LDP into our proposed \model under different noise scales $\delta$.}
\scriptsize
\begin{tabular}{cccccccc}
\hline
\textbf{Dataset} & \textbf{Noise scale} & \textbf{$\bm \delta$=0} & \textbf{$\bm \delta$=1} & \textbf{$\bm \delta$=2} & \textbf{$\bm \delta$=3} & \textbf{$\bm \delta$=4} & \textbf{$\bm \delta$=5} \\
\hline
\multirow{2}{*}{\textbf{Beauty}} & HR@50 & \bm{$0.2024$} & $0.1981$ & $0.1798$ & $0.1726$ & $0.1631$ & $0.1566$ \\
& NDCG@50 & \bm{$0.0730$} & $0.0633$ & $0.0591$ & $0.0587$ & $0.0496$ & $0.0443$ \\
\hline
\multirow{2}{*}{\textbf{TH}} & HR@50 & \bm{$0.2269$} & $0.2025$ & $0.1708$ & $0.1660$ & $0.1550$ & $0.1529$ \\
& NDCG@50 & \bm{$0.0612$} & $0.0538$ & $0.0442$ & $0.0394$ & $0.0371$ & $0.0334$ \\
\hline
\multirow{2}{*}{\textbf{TG}} & HR@50 & \bm{$0.2979$} & $0.2897$ & $0.2861$ & $0.2841$ & $0.2814$ & $0.2775$ \\
& NDCG@50 & \bm{$0.2613$} & $0.2471$ & $0.2402$ & $0.2393$ & $0.2385$ & $0.2365$ \\
\hline
\multirow{2}{*}{\textbf{DM}} & HR@50 & \bm{$0.2943$} & $0.2786$ & $0.2668$ & $0.2650$ & $0.2603$ & $0.2551$ \\
& NDCG@50 & \bm{$0.2820$} & $0.2594$ & $0.2489$ & $0.2431$ & $0.2394$ & $0.2263$ \\
\hline
\multirow{2}{*}{\textbf{OP}} & HR@50 & \bm{$0.0614$} & $0.0119$ & $0.0109$ & $0.0105$ & $0.0100$ & $0.0095$ \\
& NDCG@50 & \bm{$0.0198$} & $0.0031$ & $0.0027$ & $0.0026$ & $0.0025$ & $0.0023$ \\
\hline
\end{tabular}
\label{tab:performance_of_ldp}
\end{table}

\section{Conclusion}
To address the limitations of existing federated recommender systems in handling rich multimodal content, we propose \model, a flexible and efficient framework that systematically integrates multimodal information into the FR pipeline. By offloading the heavy multimodal representation learning to the server side and equipping clients with compact, distilled representations, \model effectively alleviates client-side computation and communication burdens. The introduction of group-aware multimodal aggregation allows for adaptive knowledge sharing among users with similar preferences, while the preference-guided distillation ensures that multimodal signals are aligned with personalized recommendation objectives. Extensive empirical studies demonstrate the consistent superiority of \model in both recommendation performance and system efficiency across multiple datasets.

\begin{acks}
Chunxu Zhang, Weipeng Zhang, Zhiheng Xue and Bo Yang are supported by the National Natural Science Foundation of China under Grant Nos. U22A2098, 62172185, 62206105 and 62202200; the Major Science and Technology Development Plan of Jilin Province under Grant No.20240212003GX, the Major Science and Technology Development Plan of Changchun under Grant No.2024WX05. Riting Xia is supported by the National Natural Science Foundation of China under Grant Nos. 62506177; the Inner Mongolia Autonomous Region Natural Science Foundation under Grant No.2025QN06010.
\end{acks}

\newpage
\bibliographystyle{ACM-Reference-Format}
\balance  
\bibliography{sample-base}

\appendix

\begin{table*}[!t]
\renewcommand{\arraystretch}{1.0}
\centering
\caption{Efficiency comparison results.}
\label{tab:efficiency}
\resizebox{\textwidth}{!}{
\begin{tabular}{c|l|c|c|c}
\toprule[1.0pt]
\textbf{Dataset} & \textbf{Model} & \textbf{Client\_Storage\_Overhead} (M) & \textbf{Communication\_Overhead} (M) & \textbf{Client\_Epoch\_Time} (ms) \\
\midrule[1.0pt]
\multirow{5}{*}{\textbf{Beauty}} & \textbf{FedMMR} & (0, 1.76) & (0, 1.23) & 7.00 \\
& \textbf{AMMFRS} & (0.36, 0.74) & (0.36, 0.74) & 3.60 \\
& \textbf{FedMR} & (0.36, 0.88) & (0.36, 0.86) & 9.10 \\
& \textbf{FedMLP} & (0.36, 0.84) & (0.36, 0.84) & 4.00 \\
& \textbf{Ours} & (0, 0.03) & (0, 0.03) & 2.60 \\
\midrule[1.0pt]
\multirow{5}{*}{\textbf{Tools\_and\_Home}} & \textbf{FedMMR} & (0, 1.78) & (0, 1.25) & 6.80 \\
& \textbf{AMMFRS} & (0.41, 0.76) & (0.41, 0.76) & 4.50 \\
& \textbf{FedMR} & (0.41, 0.90) & (0.41, 0.88) & 9.05 \\
& \textbf{FedMLP} & (0.41, 0.86) & (0.41, 0.86) & 4.20 \\
& \textbf{Ours} & (0, 0.03) & (0, 0.03) & 2.30 \\
\midrule[1.0pt]
\multirow{5}{*}{\textbf{Toys\_and\_Games}} & \textbf{FedMMR} & (0, 2.29) & (0, 1.77) & 6.90 \\
& \textbf{AMMFRS} & (1.44, 1.28) & (1.44, 1.28) & 3.30 \\
& \textbf{FedMR} & (1.44, 1.41) & (1.44, 1.40) & 8.65 \\
& \textbf{FedMLP} & (1.44, 1.38) & (1.44, 1.37) & 3.70 \\
& \textbf{Ours} & (0, 0.10) & (0, 0.10) & 2.30 \\
\midrule[1.0pt]
\multirow{5}{*}{\textbf{Digital\_Music}} & \textbf{FedMMR} & (0, 2.76) & (0, 2.23) & 7.50 \\
& \textbf{AMMFRS} & (2.36, 1.74) & (2.36, 1.74) & 3.95 \\
& \textbf{FedMR} & (2.36, 1.87) & (2.36, 1.86) & 9.40 \\
& \textbf{FedMLP} & (2.36, 1.84) & (2.36, 1.84) & 4.40 \\
& \textbf{Ours} & (0, 0.16) & (0, 0.16) & 2.50 \\
\midrule[1.0pt]
\multirow{5}{*}{\textbf{Office\_Products}} & \textbf{FedMMR} & (0, 3.86) & (0, 3.33) & 7.50 \\
& \textbf{AMMFRS} & (4.57, 2.84) & (4.57, 2.84) & 5.00 \\
& \textbf{FedMR} & (4.57, 2.98) & (4.57, 2.97) & 8.30 \\
& \textbf{FedMLP} & (4.57, 2.94) & (4.57, 2.94) & 5.30 \\
& \textbf{Ours} & (0, 0.29) & (0, 0.30) & 3.50 \\
\bottomrule[1.0pt]
\end{tabular}
}
\end{table*}

\section{Experimental Setup}
\subsection{Dataset Summary}\label{app:dataset}
These datasets, spanning diverse domains, sparsity levels, and scales, provide users' interaction records as well as item metadata (e.g., titles and images) that constitute the raw multimodal signals for our study. Following the data preprocessing strategy in \cite{zhang2023dual}, we exclude users and items with fewer than 10 interactions. For the dataset split, we adopt the prevalent leave-one-out evaluation setting~\cite{he2017neural,zhang2023dual,li2024federated}. The statistics are summarized in Table \ref{tab:datasets}.

\subsection{Baseline Details}\label{app:baseline}
We first introduce the three representative ID-based FR backbone models. Specifically, FedNCF is one of the most representative FR models, it deploys a neural collaborative filtering recommendation model~\cite{he2017neural} on each client and learns the shared item embedding and prediction function for all clients. PFedRec\cite{zhang2023dual} and FedRAP\cite{li2024federated} are two state-of-the-art FR models that advocate learning personalized item embeddings for each client to better capture individual user preferences. Then, we give a brief summary of three existing multimodal-based FR baselines. FedMMR~\cite{li2024towards} introduces multimodal information into FR by incorporating a server-side step that aligns various modality representations with global item embeddings. AMMFRS~\cite{feng2024robust} and FedMR~\cite{li2024dynamic} enhance local item representation by integrating multimodal features, using self-attention and mixture-of-experts mechanisms, respectively.

\subsection{Implementation details}\label{app:implementation}
 Specifically, we predict over the full item set and calculate the metrics based on the rank of the test item within the top-50 recommendation list. We report the average results over five independent runs. All experiments are implemented in PyTorch and conducted on a machine with three NVIDIA GeForce RTX 3090 GPUs. For multimodal representation learning, we employ CLIP based on the Vision Transformer Base (ViT-B/32) architecture~\cite{radford2021learning,dosovitskiy2020image}. Missing visual features are imputed with the average values from available items, while missing titles are replaced with a placeholder string "There is no title." To ensure a fair comparison, we set the latent dimension to 512, the batch size to 256, and the number of communication rounds to 100, which we empirically find sufficient for convergence. We additionally conduct an exhaustive hyperparameter search for all baseline methods, grid-searching the learning rate over \{0.1, 0.01, 0.001, 0.0001\} and the weight decay over \{1e-3, 1e-4, 1e-5\}. For other method-specific hyperparameters, we either retain the values from their original publications or optimize them on our validation set until performance plateaued. Consistent with prior work \cite{he2017neural,zhang2023dual}, we adopt a negative sampling strategy with four negative instances per positive instance during training. 

\begin{table}[!t]
\centering  
    \caption{Dataset statistics.}
    \label{tab:datasets}
    \resizebox{.95\linewidth}{!}{
    \begin{tabular}{lcccc}
        \hline
        Dataset & \#Users & \#Items & \#Interactions & Sparsity \\ \hline
        Beauty      & 253     & 356     & 2,535      & 97.19\%  \\
        Tools\_and\_Home   & 242   & 398    & 3,606     & 96.26\%  \\
        Toys\_and\_Games    & 1,146   & 1,403    & 23,820     & 98.52\%  \\
        Digital\_Music   & 1,517   & 2,306   & 35,395    & 98.99\%  \\
        Office\_Products   & 6,338   & 4,461  & 97,227    & 99.65\%  \\
        \hline
    \end{tabular}
    }
\end{table}

\section{Additional Experiments}
\subsection{Efficiency Comparison Results}\label{app:efficiency}
In real-world deployments of FR systems, efficiency is as critical as performance, particularly given the limited storage and computation resources on client devices and the frequent communication required between clients and the server during federated optimization. The efficiency comparison results are summarized in Table~\ref{tab:efficiency}. The \textbf{Client\_Storage\_Overhead} column includes two metrics: (1) the size of multimodal representations stored on the client, and (2) the parameter count of the local recommendation model. The \textbf{Communication\_Overhead}  column reports: (1) the initial distribution cost of the multimodal representations, and (2) the per-round transmission cost between clients and the server. The \textbf{Client\_Epoch\_Time} column denotes the per-epoch training time of the client-side recommendation model.

For baselines AMMFRS, FedMR, and FedMLP, storing item multimodal features locally and designing complex fusion modules increase client storage burden; FedMMR’s storage overhead stems from its architecture integrating an elaborate contrastive loss. The complexity of these local models lengthens client training and enlarges per-round transmission cost between clients and server. Additionly, AMMFRS, FedMR, and FedMLP need to transmit full initial multimodal representations to all clients before training, incurring substantial communication overhead. In contrast, our method keeps item multimodal content on the server and employs a high-capacity encoder to generate representations, alleviating client computational and storage burden; only low-dimensional group-preference vectors (e.g., 2 vs. 512) are exchanged, introducing minimal communication overhead. Consequently, our method demonstrates superior efficiency, making it a more practical solution for real-world deployment under resource constraints.

\begin{figure}[htbp]
        \centering
        \includegraphics[width=\linewidth]{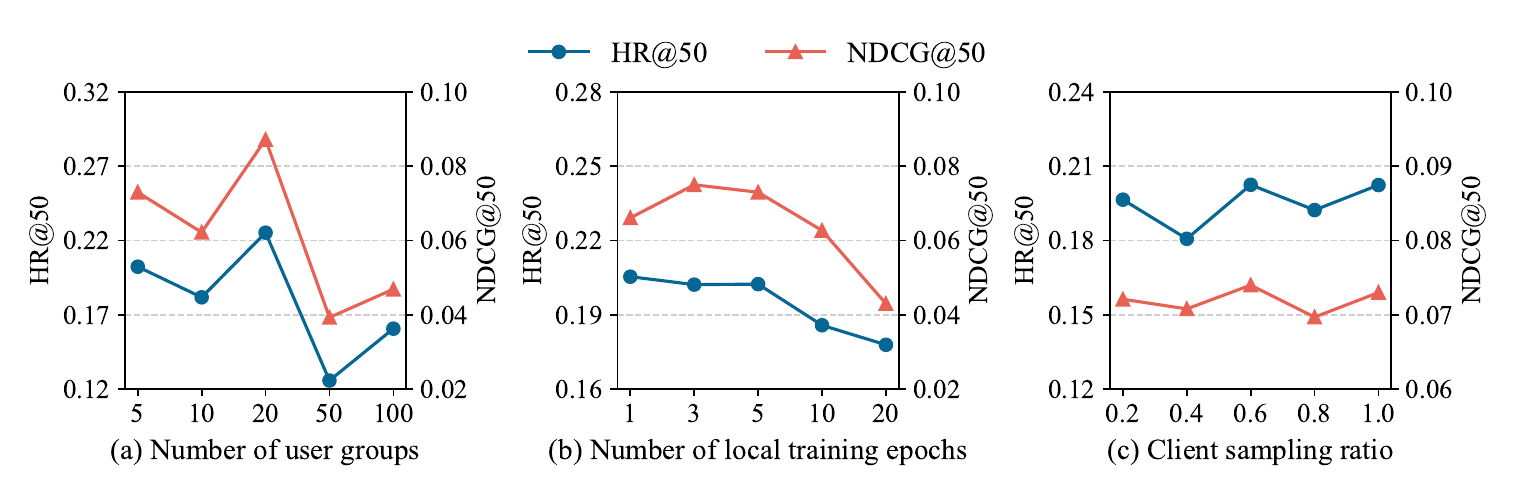}
        \caption{Hyper-parameter analysis results on dataset Beauty.}
        \label{fig:hyper_ab}
\end{figure}
  
\begin{figure}[htbp]
        \centering
        \includegraphics[width=\linewidth]{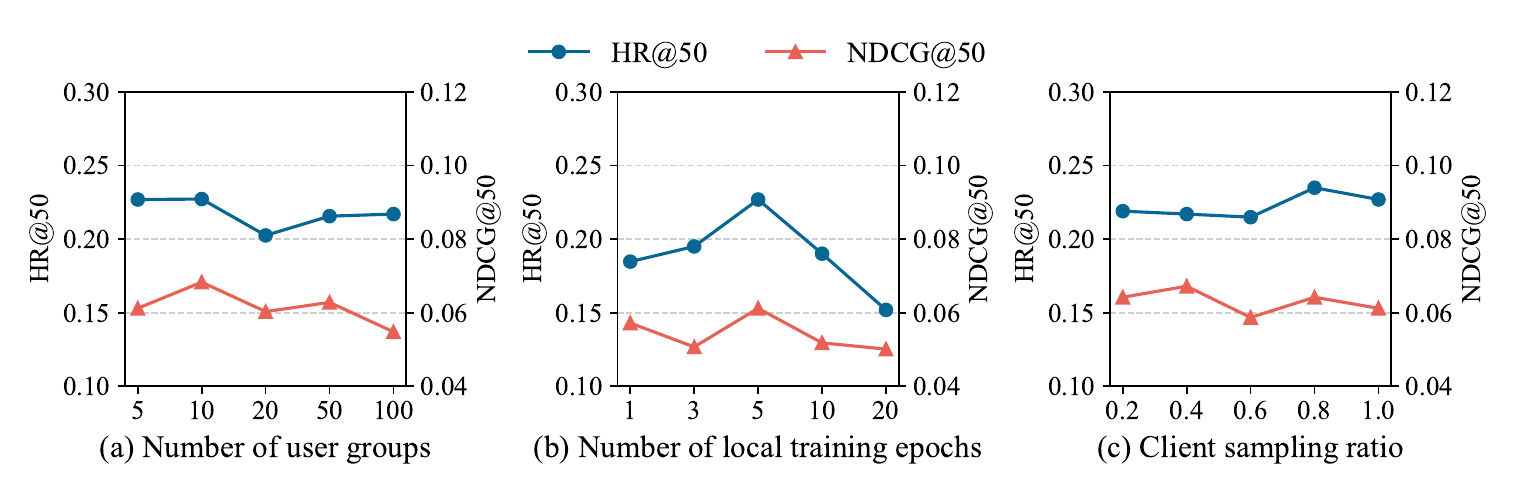}
        \caption{Hyper-parameter analysis results on dataset Tools\_and\_Home}
        \label{fig:hyper_th}
\end{figure}

\begin{figure}[htbp]
    \centering
    \includegraphics[width=\linewidth]{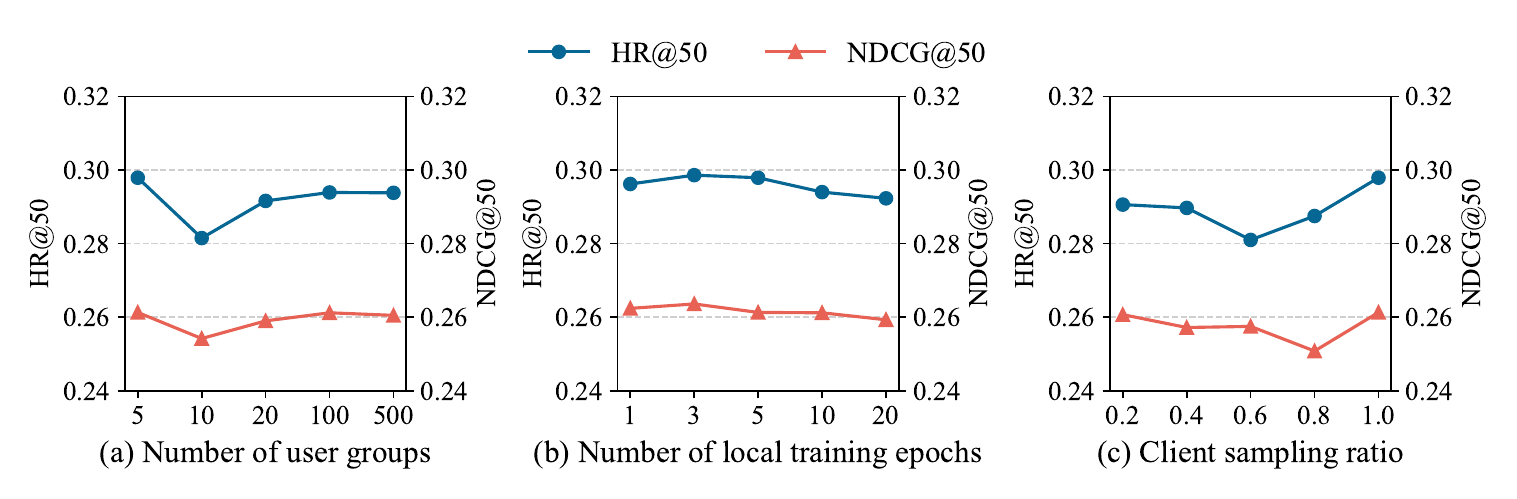}
    \caption{Hyper-parameter analysis results on dataset Toys\_and\_Game.}
    \label{fig:hyper_tg}
\end{figure}

\begin{figure}[htbp]
    \centering
    \includegraphics[width=\linewidth]{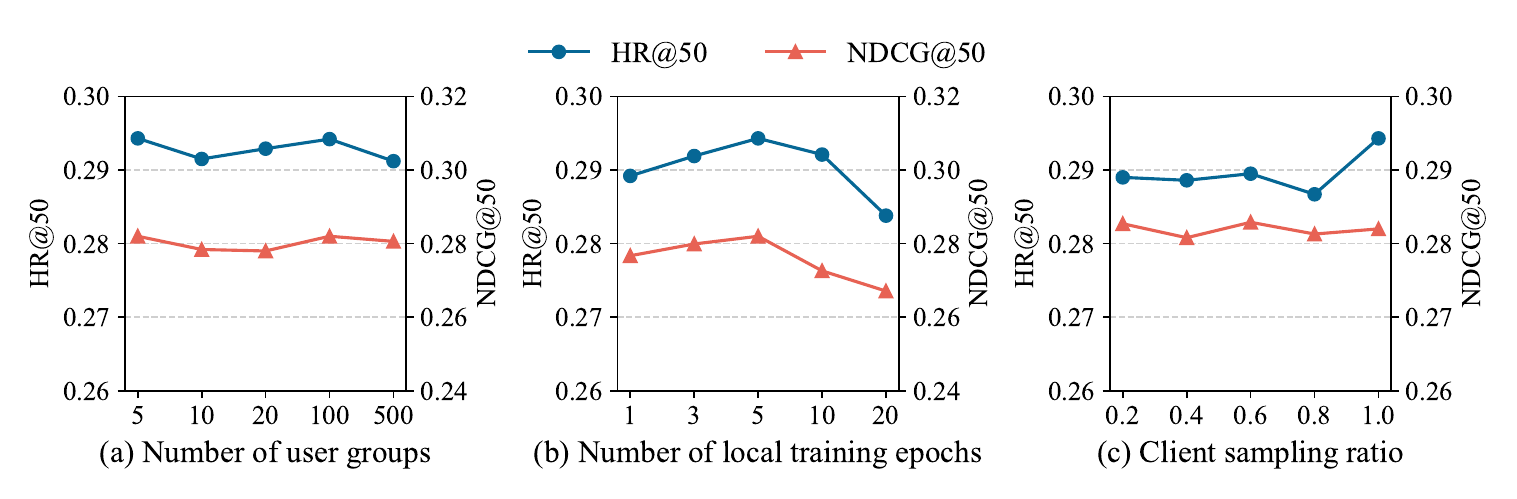}
    \caption{Hyper-parameter analysis results on dataset Digital\_Music}
    \label{fig:hyper_dm}
\end{figure}

\begin{figure}[htbp]
    \centering
    \includegraphics[width=\linewidth]{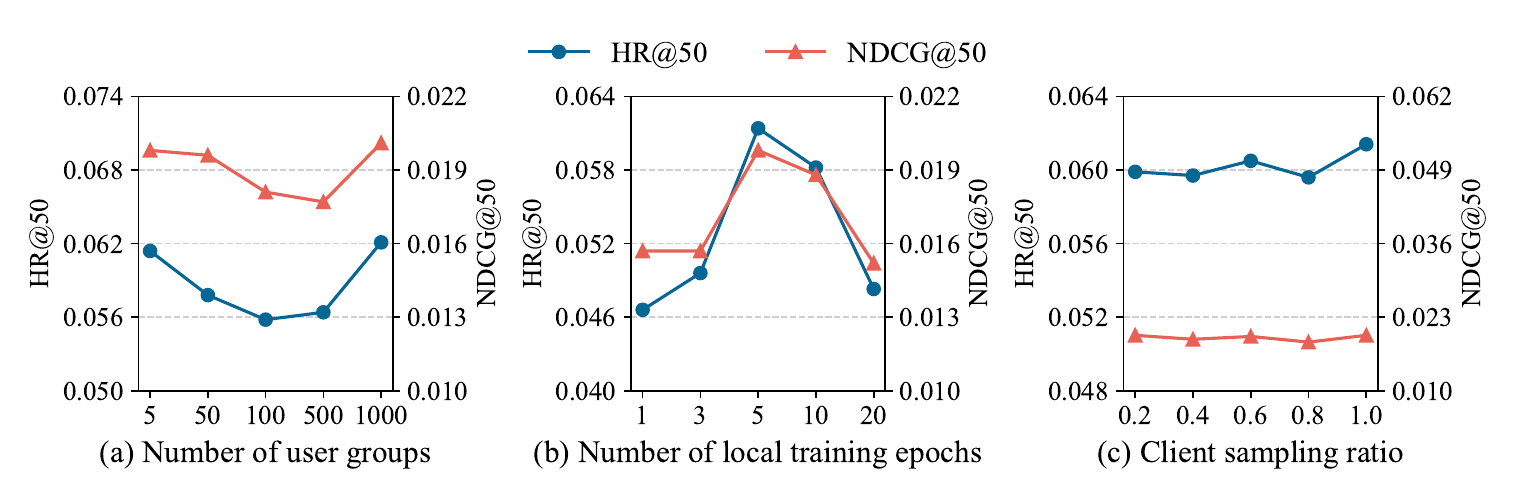}
    \caption{Hyper-parameter analysis results on dataset Office\_Products}
    \label{fig:hyper_op}
\end{figure}

\subsection{Further Hyper-Parameter Analysis Results}\label{app:hyperparameter}

\paragraph{Number of user groups $\bm{g}$.}
Based on the user scale of each dataset, we vary the number of user groups to investigate its impact on recommendation performance. Specifically, we test group numbers of $\{5, 10, 20, 50, 100\}$ for Beauty and Tools\_and\_Home, $\{5, 10, 20, 100, 500\}$ for Toys\_and\_Games and Digital\_Music, and $\{5, 50, 100, 500, 1000\}$ for Office\_Products. As shown in Figure~\ref{fig:hyper_ab}, \ref{fig:hyper_th}, \ref{fig:hyper_tg}, \ref{fig:hyper_dm}, \ref{fig:hyper_op} (a), experimental results across five datasets demonstrate that the number of user groups significantly influences performance, with the optimal value strongly depending on dataset-specific characteristics. This finding suggests that our method effectively adapts to the unique properties of each dataset by partitioning users into groups, thereby capturing diverse item preferences across these groups.

\paragraph{Number of local training epochs $\bm{H}$.}
We evaluate the impact of the number of local training epochs from the set $\{1, 3, 5, 10, 20\}$ across all datasets. As shown in Figure~\ref{fig:hyper_ab}, \ref{fig:hyper_th}, \ref{fig:hyper_tg}, \ref{fig:hyper_dm}, \ref{fig:hyper_op} (b), experimental results indicate that setting the number of local epochs to 5 yields optimal or near-optimal performance in all cases. Too few local epochs result in underfitted and noisy updates, whereas an excessive number induces overfitting to local data and escalates computational costs, ultimately degrading model performance.

\paragraph{Client sampling ratio $\bm{\beta}$.}
We examine the impact of the client sampling ratio across all datasets by testing values in $\{0.2, 0.4, 0.6, 0.8, 1.0\}$. As shown in Figure~\ref{fig:hyper_ab}, \ref{fig:hyper_th}, \ref{fig:hyper_tg}, \ref{fig:hyper_dm}, \ref{fig:hyper_op}(c), different sampling ratios generally yield comparable final performance. While a lower sampling ratio involves fewer clients per communication round, this can be offset by increasing the number of rounds to achieve convergence. In contrast, higher sampling ratios accelerate convergence and improve computational efficiency.

\end{document}